\documentclass[aps,epsf,epsfig,preprint]{revtex4}
\usepackage{graphicx} \usepackage{amssymb,amsmath, amsthm}
\newcommand{\beq}{\begin{equation}} 
\newcommand{\eeq}{\end{equation}}
\newcommand{\barray}{\begin{eqnarray}}
\newcommand{\earray}{\end{eqnarray}}

\begin{document}

%\preprint{}

\title{
The Hall Number, Optical Sum Rule and Carrier Density for the $t$-$t'$-$J$ model
}


\author{Jan O. Haerter and 
 B. Sriram Shastry}
\address{Physics Department, University of California,  Santa Cruz, Ca 95064 } 
\date{October 29, 2007}

\begin{abstract}
We revisit  the relationship between three classical measures of particle number, namely the chemical doping $x$, the Hall number $x_{hall}$ and the particle number inferred from the optical sum rule $x_{opt}$. We study the $t$-$t'$-$J$ model of correlations on a  square lattice, as a minimal model for High $T_c$  systems, using  numerical methods to evaluate the low temperature Kubo conductivites.   These  measures  disagree  significantly in this type of system, owing to Mott Hubbard correlations.
The Hall constant has a complex behavior with several changes of sign as a function of filling $x$,  depending upon the model parameters.
Thus  $x_{hall}$  depends sensitively on $t'$ and $J$, due to a kind of quantum interference.  
\end{abstract}

 \maketitle
%\keywords{}


\section{Introduction}

The traditional strategy, of converting a measured Hall constant or an optical sum rule  to an electron count, runs into serious difficulties when interactions are strong within a  lattice fermi system. The non conservation of the lattice current, unlike its continuum counterpart, changes the f-sum rule drastically to involve non universal variables such as the kinetic energy expectation. Similarly the Hall constant suffers serious many body renormalization due to physics associated with the Mott Hubbard correlations; holes in the Mott Insulator have little resemblance to carriers in uncorrelated bands.
  
This problem has very recently been revived in the context of LSCO ($La_{2-x}Sr_xCuO_4$) \cite{tsukada}, the authors refining  the initial work of Takagi et.al. \cite{hwang} using high quality thin films. This is a particularly suitable system since the doping can  be tuned all the way from the lightly doped to  the overdoped Fermi liquid regime. The Hall constant provides  several outstanding puzzles, firstly  a change of  sign from $R_H >0$ at small $x$ to $R_H<0$ at  $x \geq .3$, where $x= 1-n $ is  the number of holes  per copper. Further there is a quite substantial T dependence for small $x\leq .3$. The problem is compounded by the angle resolved photoemission (ARPES) data\cite{yoshida}, which shows that the topology of the fermi surface remains electron like from $x\geq .18$, so that the change of sign cannot be easily ascribed to a fermi surface distortion.  There is a notable  recent attempt\cite{narduzzo} to rationalize the observed behavior using theoretical ideas\cite{varma} invoking strong and anisotropic impurity (elastic) scattering. Thus  factors extrinsic to the two dimensional plane are invoked to understand the $x, T$ dependence.

As noted recently \cite{tsukada}, the measured Hall constant in better samples continues  towards large negative values as $x\rightarrow 1$, in contrast to the early data\cite{hwang} that appeared to saturate. This overall behavior of the Hall constant, namely a 
large positive value as $x \rightarrow 0$, and a large negative value as $x \rightarrow 1$ are precisely of the kind {\em intrinsic to  a Mott Hubbard system}, as first pointed out in \cite{sss}.
Thus a final theory would reconcile impurity scattering to intrinsic factors of the kind we study in this work.  The early work of Ref{\cite{sss} (Shastry-Shraiman-Singh (SSS)] showed that the high frequency Hall constant shows  a sensitivity to half filling and hence to Mott Hubbard physics. It  gives a divergent Hall constant at half filling, together with (at least) three zero crossings,  as the band filling $n ={N_{electrons}}/{N_{sites}} $ varies from 0 to 2. Other recent ideas\cite{phillips} on Mott physics lead to comparable results. The results of SSS were obtained for  a nearest neighbor $t-J$ model on the square lattice at high temperature. This  led to a hole like Hall constant for $0 \leq x \leq .3$, followed at large $x$ by an electronic Hall constant. This change of sign is in agreement with the experiments\cite{tsukada, hwang} on LSCO, but not so with several other High $T_c$ compounds (e.g. $YBa_2Cu_3O_{6+\delta}$)  that do not show  a change of sign within the  available range of doping. Thus the problem of understanding the Hall constant in the various  classes of High $T_c$ systems remained  unresolved, a task that we return to  in this work.

 Further  recent experimental work of Refs\cite{balakirev,basov} on the optical mass and anamolous behavior of the Hall number in
good samples of LSCO adds 
 motivation to  this effort. Here  we address the 
problem of computing the effect of correlations on the  effective carrier count, or equivalently the Hall constant and the optical mass, for a model system, the $t$-$t'$-$J$ model on a square lattice.
While the optical mass is quite straightforward to address, using exact computation of the expectation value of the ``stress tensor'' or kinetic energy, the case of the Hall constant is quite non trivial, as elaborated below. 

A study  of the fermi surface is another possible source of information on the Hall constant. 
We have alluded to the recent work in Ref\cite{yoshida} on the ARPES derived shape of the fermi surface for LSCO at all dopings. Theoretically  however, this is  a vexed issue.  Firstly, in an interesting numerical study, the Luttinger theorem's validity in  t-J models describing strongly correlated matter  has recently been questioned\cite{prelovsek}. Even when the theorem   does apply, the possibility of shape deformation\cite{nozieres} is strong. The implications are that for  any choice of bare band parameters $t,t'$ made,  leading to ansiotropic {\em bare  fermi surfaces}, one must excercise caution in interpreting the {\em observed fermi surface}. This is so, since the fermi surface is further deformed in an area (volume) preserving fashion due to the interactions, leading to the experimentally observed  {\em renormalized fermi surface}. The final observed  fermi surface is expected to be quite different from the starting shape, since there are  reasons to expect a strongly momentum dependent self energy\cite{momentum}. There have been few studies of this difficult issue in literature, since it requires the knowledge of the momentum dependence of the self energy. The situation for numerical studies is also rather  unfavourable, since very few values of the momentum are available in finite sized clusters, making it hard to determine a surface. We therefore avoid any discussion of it, and continue with a study of objects that are more direct for the Hall constant.

Motivated by  the triangular lattice system $Na_xCoO_2$,
 we\cite{curie_weiss} have recently studied the Hall constant as a function of temperature as well as frequency quite thoroughly. We  used exact diagonalization to compute the 
{\em exact Kubo formulas} for the Hall constant, to benchmark  the high frequency approximations to the same.
This parallel experience is helpful in the present context. Quite encouraging is the result that the frequency dependence is mild in essentially all cases studied, 
so that one can get a reasonable  estimate of the Hall constant from the high frequency results. The temperature dependence of the Hall constant is serious for the triangular lattice, owing to  the peculiar structure of the closed loops on the former\cite{kumar,ong}.  For the square lattice
this is not expected to be  as serious, on general  grounds.  It is however found   that  the underdoped cases {\em do have} an inexplicable T sensitivity\cite{balakirev,narduzzo,hwang}  especially at low T and low $x$, although the scale of the $T$ dependence is modest in comparision to that in the triangular lattice cobaltates. We are unable to address this issue here, as it seems to be related to  the essential complexity of the pseudogap phase. Our computations are all at a low (effectively zero) temperature.

In this work, we go beyond the framework of SSS, by studying the
 $t-t'-J$ model Eq(\ref{Ham}) on the square lattice (without the restriction of high temperature expansions). 
    This  is  an often used  model to describe the physics of the copper oxide planes in High $T_c$ cuprates. The addition of the second neighbor hopping $t'$ is required by LDA calculations \cite{andersen,bansil} in order to fit to an effective tight binding  model. Most importantly for our purpose, it extends nontrivially the simple $t-J$ model studied earlier and yields a rich variety of behavior of the Hall constant that seems to have the potential to explain  the observed  experimental diversity. In this work, we present preliminary results in this direction, by computing the Hall constant on small clusters of the above model for various values of the ratio $ t'/|t|$ for small clusters of up to 15 sites.   We demarcate regions where the change of sign is observed as in SSS,  from those where apparently no change occurs. Morover, we provide rough estimates of the effective number of holes as a function of the chemical doping and the ratio $t'/|t|$. We are unable to examine more subtle issues such as the possible existence of a quantum critical point in Ref\cite{balakirev}, but rather wish to provide a rough base line from which one can build a more elaborate theory.

Given a theoretical model with $x$ holes per copper, one can compute an effective doping $x_{hall}$ from the Hall constant $R_H$ via
$ x_{hall}\equiv v/(R_H q_e)  $, where  $v$ is the volume per copper and  
  $q_e= - |e|$ the elementary unit of charge. Similarly, given  the optical conductivity $\sigma(\omega)$, we can define an optical doping $x_{opt}$. Consider the f-sum rule\cite{plasmasumrule,shastry_sum_rule} on a lattice
\beq \int_0^\infty \Re e \ \sigma(\omega) \ d\omega= \frac{\pi}{2\hbar{\mathcal L}}\langle \tau^{xx}\rangle \equiv \frac{ q_e^2 \pi x_{opt}}{ 2 m_b v},\label{f_sum_rule}\eeq 
with  ${\mathcal L}$ the crystal volume, $m_b$ the band electron mass (defined below),  and 
\beq\tau^{xx}= \frac{q_e^2}{\hbar}\sum_{k,\sigma} \frac{d^2 \varepsilon(k)}{d k_x^2}  c_\sigma^\dagger(k) c_\sigma(k),\eeq
 the stress tensor. We can define the effective  plasma frequency from $ \omega_p= \sqrt{\frac{4 \pi q_e^2 x_{opt}}{m_b v}}$ so that the f-sum rule leads to $\omega_p^2/8$ as usual.
In the case of a parabolic noninteracting band this object reduces to the familiar 
result, and provides a natural generalization to the tight binding cases.

We note that the optical sum rule can be also interpreted as a renormalization of the effective mass, since it  only measures the ratio of filling to mass. We favor the above factorization, wherein the $x_{opt}$ contains {\em all the many body renormalizations}, but not  the band effects.
The band effects can be absorbed into the (optical) band mass $m_b$ meaningfully as follows.  We define $m_b\equiv m_b(t,t',x)$, where 
\beq\frac{\hbar^2 n}{ m_b v} = \frac{1}{{\cal L}} \sum_{k,\sigma} \frac{d^2 \varepsilon(k)}{d k_x^2}  <c_\sigma^\dagger(k) c_\sigma(k)>_{0},\label{x_optical}\eeq
with $n=1-x$ the electron density  per copper    with the average being carried out in the noninteracting band. The  ratio $m_b/m_e$ as a function of its various arguments is easily evaluated, where $m_e$ is the bare electronic mass. For the case of $t'=0$ and $t=5160^0K$, the  ratio $m_b/m_e \sim 1.0 $. In view of this close proximity between the band and bare masses, we simply put $m_b=m_e$. The lattice parameter used in our computations is $a_0=3.79\;10^{-10}\;m$ appropriate to LSCO, so the use of our results for other materials would require a small adjustment factor for the atomic volume.

In comparing with experiments, it must also  be borne in mind that the projected t-J model contains only a part of the spectral weight, since it describes the low energy part of the Hilbert space. Literally it implies that the charge transfer gap is sent to infinity, so the integration in the sum rule must be cut off at roughly some fraction of the charge transfer gap. In practice \cite{basov},  the upper limit for the frequency integral is often chosen precisely in such a way so that a comparision is not unjustified.

As stated above, in a weakly correlated  system  all three particle numbers are expected to be equal, hence $x=x_{hall}=x_{opt}$. In strongly-correlated systems, however, it is expected  that this simple relation no longer holds since different variables undergo different renormalizations. Further these many body effects also depend upon the initial starting model parameters non trivially, including the band structure effects.  For the band structure in cuprate materials, several groups have emphasized the need to include second and possible further neighbour hoppings\cite{andersen,bansil}. In the following, we attempt to shed light on the different many body effects for a given  chemical doping and for different ``band parameters" $t,t'$ as well as $J$.

In section II, we state the model and state the formulas that are computed as well as some indication of the methods used. In section III, we discuss the results for the Hall constant, its frequency dependence, the effective Hall number  and the optical mass.
In section IV, we make concluding remarks.

\section{Model and exact diagonalization}
We study  the $t$-$t'$-$J$ model on the  square lattice, as a model for  the strongly-correlated cuprates.
\begin{widetext}
 \beq \hat{H}=-t\sum_{<i,j>,\sigma}
\hat{P}_G \hat{c}_{i \sigma}^{\dagger}\hat{c}_{j
\sigma}\hat{P}_G-t'\sum_{\ll i,j\gg ,\sigma}
\hat{P}_G \hat{c}_{i \sigma}^{\dagger}\hat{c}_{j
\sigma}\hat{P}_G+J\sum_{<i,j>}\left(\hat{S}_i \cdot
\hat{S}_j-\frac{\hat{n}_i \hat{n}_j}{4}\right)\label{Ham}\eeq
\end{widetext}
 where $ \hat{c}_{i
\sigma}^{\dagger}$ ($\hat{c}_{i \sigma}$) creates (annihilates) an
electron of spin $\sigma$, $\hat{S}_i$ is the three-component
spin-operator, $\hat{n}_i$ is the number operator and $i$ specifies
the lattice site. $\hat{P}_G$ denotes the Gutzwiller projector and the
summation is over all nearest (second nearest) neighbor pairs $<i,j>$ ($\ll i,j\gg $). Here, $t$ ($t'$) is the nearest (second nearest) neighbor hopping amplitude.
\begin{center}
\begin{figure}[h]
\includegraphics[width=9cm]{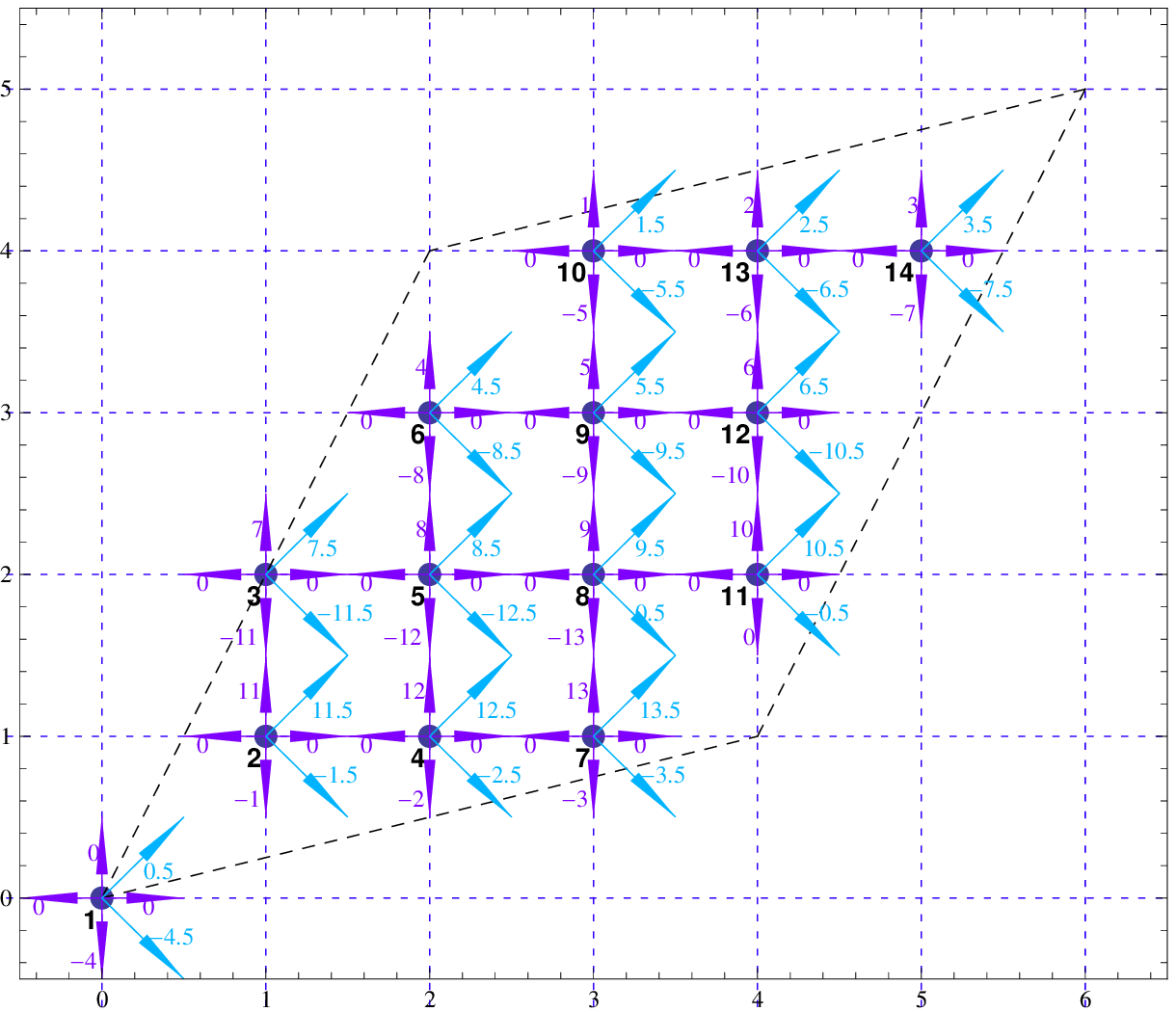}\\
\includegraphics[width=9cm]{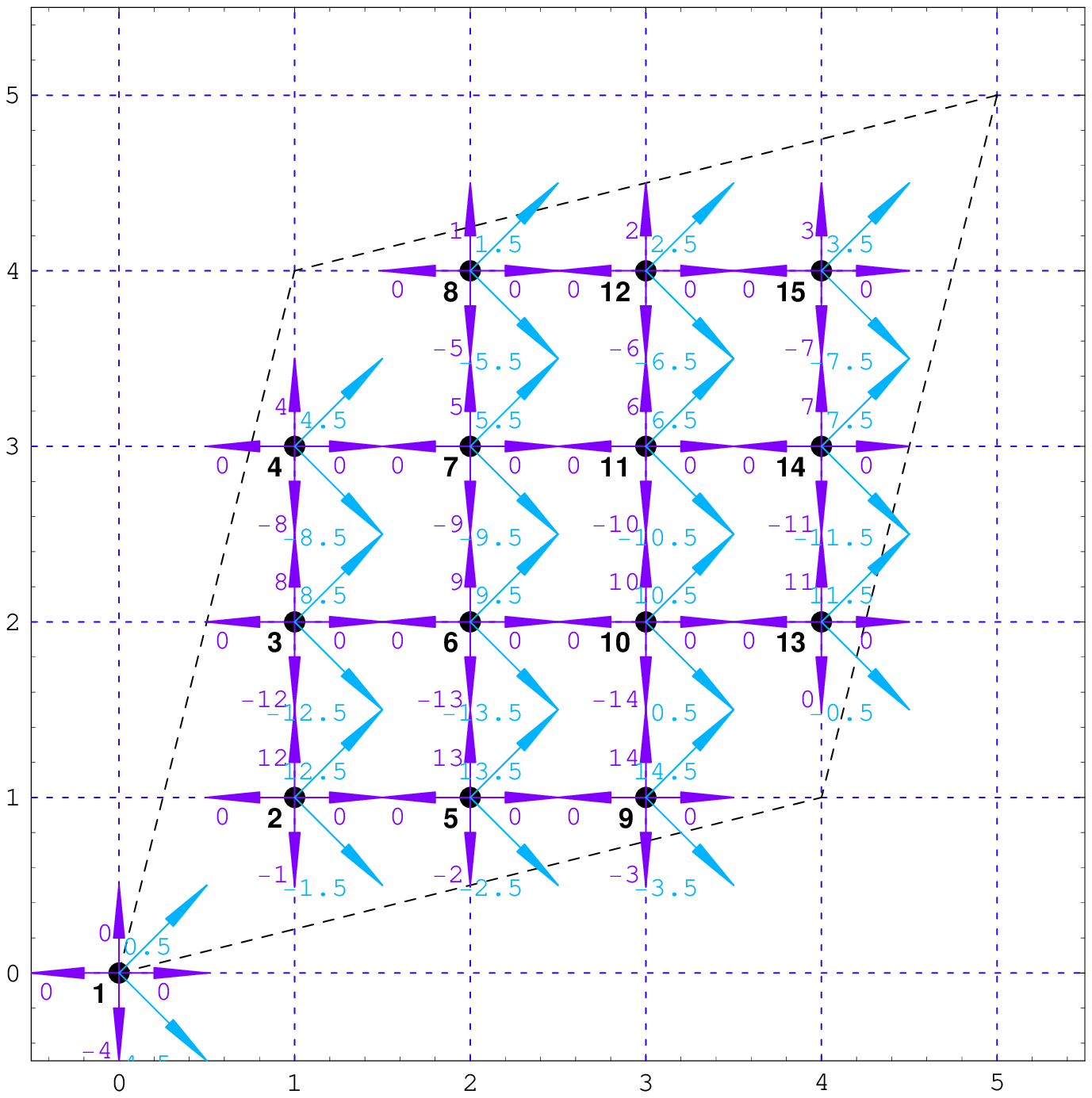}
  \caption{(color on-line) Image shows the finite clusters used in our computations, arrows indicate Peierls phase-factors with second neighbor hopping in magnetic field. Top (bottom) panel shows the 14(15) site square lattice cluster.}
\label{14_site_square}
\end{figure}
\end{center}
We have employed toroidal geometries with $L=$ $14$ and $15$  sites. Whenever possible, we reduce the computational effort by exploitation of space group symmetries. The symmetries are lowered  upon
the introduction of a magnetic field, as relevant for the evaluation
of the Hall coefficient and the Hall number. For example simple translations are no longer good symmetries.
 The magnetic field is introduced by the usual
Peierls substitution which modifies the hopping $t$ between sites $i$
and $j$ by  \beq\label{peierls} t\rightarrow t_{ij}({\bf A})=t\exp
\left( i\frac{2\pi}{\phi_0}\int_i^j {\bf A}\cdot d{\bf
s}\right)\;,\eeq where ${\bf A}$ is the magnetic vector potential and
$\phi_0= \frac{h c}{|q_e|}$ the flux quantum. We define the dimensionless flux threading a square
plaquette as $\alpha\equiv \frac{2\pi}{\phi_0}\oint
{\bf A}\cdot d{\bf s}$. In finite systems, the value of the smallest
non-zero magnetic field is limited to values of $\alpha\geq \pi/l$
where $l$ is the length of a periodic loop in the system. Through a
particular gauge we can achieve $l$ equal to the number of square
faces in the cluster, this guarantees the equality of the flux values
through all plaquettes. In the case of second neighbor hopping we introduce additional phase factors along the diagonals of the square plaquettes in such a way that all fluxes through the resulting triangular plaquettes become equal.
An example is given in
Fig. \ref{14_site_square}. A similar strategy has been followed in
the case of the square lattice quantum Hall effect\cite{kohmoto}. The Hall-coefficient has been investigated earlier  within the nearest-neighbor Hubbard and $t$-$J$ models\cite{sss,phillips,jaklic, assaad}. In this work, we are most interested in the dependence of this quantity on a second-neighbor hopping parameter. This task  seems necessary to include in the starting model to explain the wide variety of behavior observed in the cuprates, and has not apparently been undertaken earlier.

\section{Results}

\subsection{Frequency-dependence of $R_H$}
To further establish the validity of the high-frequency limit $R_H^*$ of the Hall-coefficient we compute explicitly the real and imaginary part of the frequency-dependence of $R_H(\omega)$ through the Kubo-formula \cite{sss,shastry_sum_rule} for the electrical conductivity $\sigma_{\alpha\beta}$ 
\begin{widetext}
\beq \sigma_{\alpha \beta}(\omega)=\frac{i}{\omega \, \Omega }\left[\langle \tau^{\alpha\beta}\rangle-\frac{1}{\mathcal{Z}}\sum_{\mu\nu}\frac{e^{-\beta\epsilon_\nu}-e^{-\beta\epsilon_\mu}}{\epsilon_\mu-\epsilon_\nu-\omega-i\eta}\langle\nu|J_\alpha|\mu\rangle\langle\mu|J_\beta|\nu\rangle\right]\;,\label{lehmann_cond}\eeq
\end{widetext}
where $\beta$ the inverse temperature, $ \Omega $ the volume of the system, $\hbar \rightarrow 1$,  and the sum is taken over all eigenstates of the system. 
The symbol $J_\alpha$ stands for the current operator in a field, and $\cal{Z}$ is the partition function.
The complex frequency dependent Hall-coefficient can then be expressed as\cite{sss}
\beq R_H= \lim_{B \rightarrow 0} \frac{\sigma_{xy}}{ B \sigma_{xx}\sigma_{yy}},\label{linear_hall}\eeq
where $B$ is the magnetic field transverse to the plane and $\sigma_{\alpha\beta}$ is the conductivity tensor as defined in Eqn. \ref{lehmann_cond}.  The transport Hall-coefficient $R_H^{Tr}\equiv \lim_{\omega\rightarrow 0} R_H(\omega)$ is connected to the imaginary part of $R_H$ by a dispersion relation following from causality.
Since $ R_H(\omega)$ is analytic in the upper half of the complex $\omega$ plane, and has a finite limit at infinite $\omega$, we may write
\beq R_H(\omega) = R_H(\infty) - \int_{-\infty}^{\infty} \frac{ d \nu}{\pi}\; \frac{ \Im m R_H(\nu)}{\omega - \nu + i 0^+}, \eeq
therefore setting $\omega=0$ we get the interesting result:
\beq {\Re e}R_H(0)= R_H^*+ \frac{2}{\pi}\int_0^\infty\frac{{\Im m} R_H(\nu)}{\nu}\;d\nu \;. \label{eqimrh}\eeq
Note that the two  regions where $R_H(\omega)$ is real,  are $\omega \rightarrow 0$ and $\omega \rightarrow \infty$. As before \cite{sss}, we define $ R_H^* \equiv  R_H(\infty)$.
This equation  quantifies the difference between the experimentally measured dc-Hall coefficient and the theoretically more accessible infinite frequency limit. The second term on the right is often found numerically to be quite small, and interestingly is  an independently measurable object. 
We are aware of few such recent such measurements of $\Im m R_H(\omega)$\cite{drew} for a  correlated system, and believe  that it is  be worth having more extensive measurements of this object. With a measurement/computation of $\Im m R_H(\omega)$, the second integral in Eq(\ref{eqimrh}) can be computed numerically with some confidence, since it involves an integration, which provides an automatic smoothing of the data.   For a few extreme  values of the chemical doping $x$ we show the real and imaginary part of $R_H$ in Fig. \ref{frequency_dependence_rh}. The case of small dopings has the largest $\omega$ correction, for larger dopings the correction seems to fall away rapidly. These computations demonstrate the typical magnitudes of the  frequency-dependence of the imaginary part of $R_H(\omega)$.  In the range $x\geq \sim .18$,  we  estimate  $R_H^*$  to be quite close to  the dc value. Thus it is enough for qualitative purposes to ignore the distinction between the two variables.  We plan to return to more extensive computations in the future, in order to extract the transport Hall constant. For the present computation of the Hall number $x_{hall}$,  with the above cautionary remark,    we use the high frequency object\cite{sss}: 
\beq R_H^*\equiv \lim_{B\rightarrow 0}\left(-\frac{i \Omega }{Bq_e^2} \frac{\langle [J_x,J_y]\rangle}{\langle\tau^{xx}\rangle^2}\right)\label{R_H_star}.\eeq 

\begin{center}
\begin{figure}[h]
\includegraphics[width=16cm]{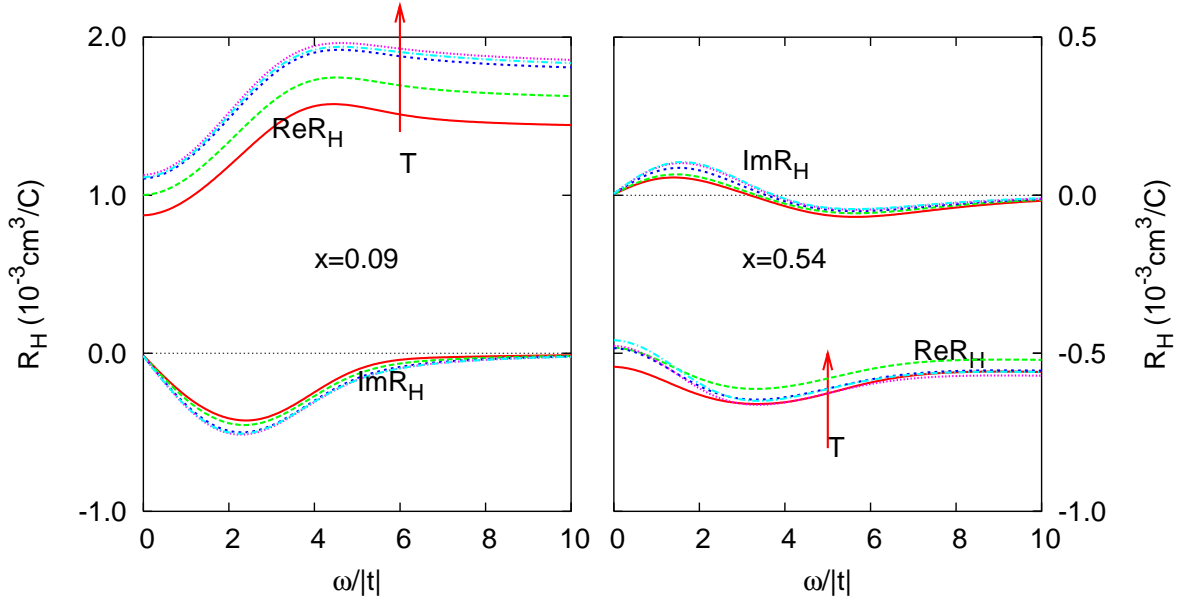}
  \caption{(color on-line) Frequency-dependence of the Hall coefficient on the simple square lattice for $x=1/11$ ({\bf l.}) and $x=6/11$ ({\bf r.}), the values for doping are chosen as extreme cases and we expect intermediate behavior of $R_H(\omega)$ in between these two values, hence an overall modest frequency-dependence. The range of $T$ is from $1.6 |t|$ to $ .2 |t| $
}
\label{frequency_dependence_rh}
\end{figure}
\end{center}

\subsection{Hall Coefficient}
We now analyse the doping-dependence of the ground state Hall coefficient $R_H^*$ when a second neighbor hopping is included in the Hamiltonian of Eqn. \ref{Ham}.
We begin with  $J=0$ (bottom panel of Fig\ref{Hall_coefficient}). We find that the value of a zero-crossing at finite doping is in fact highly sensitive to the value of $t'$ (Fig. \ref{Hall_coefficient}). At $t'=0$ the computations show a zero-crossing near $x=1/3$ similar to the prediction from the high-temperature expansion\cite{sss}. Turning on a positive $t'$, the zero crossing is pushed to lower $x$ and is essentially invisible in our studies, since we cannot reach appreciably below $x=.12$. Turning on a negative $t'$, the zero crossing is more pronounced and is pushed out to lager $x$! In order to place these results in context, recall that a
positive $t'$ for hole doping leads to electronic frustration, as in the triangular lattice sodium cobaltate\cite{curie_weiss}. A negative $t'$, on the  on the hand, causes a ferromagnetic Nagaoka tendencey (towards a large fermi surface). While quantum fluctuations  as well as the pernicious influence of the exchange constant J prevent the collapse into ordered states, these tendencies do seem to influence the behavior of the Hall constant.  We thus interpret the strong dependence on the sign of  $t'$ as a quantum  interference effect. In our earlier study on the triangular lattice \cite{curie_weiss},  we found results that are very similar to  what we find here for $t'>0$.

To study the effect of $J>0$, we compute the Hall constant at two representative values of $J$ (two upper panels of Fig\ref{Hall_coefficient}). We see that exchange has a similar effect to $t'>0$, both lead to a suppression of the magnitude of the Hall constant. The influence on the zero crossing is more complex: in some cases it is suppressed (in our computationally available range of $x$), in others we find an extra zero crossing at lower $x$ where $R_H$ becomes negative. Presumably at very small $x$ it turns positive and diverges due to the Mott Hubbard gap. This would imply that the Hall constant has a total of {\em three}  zero crossings in the range $0\leq x \leq 1$ ( or six in the range $0\leq n\leq 2$), in contrast to a single crossing for the uncorrelated case. 
\begin{center}
\begin{figure}[h]
\includegraphics[width=11cm]{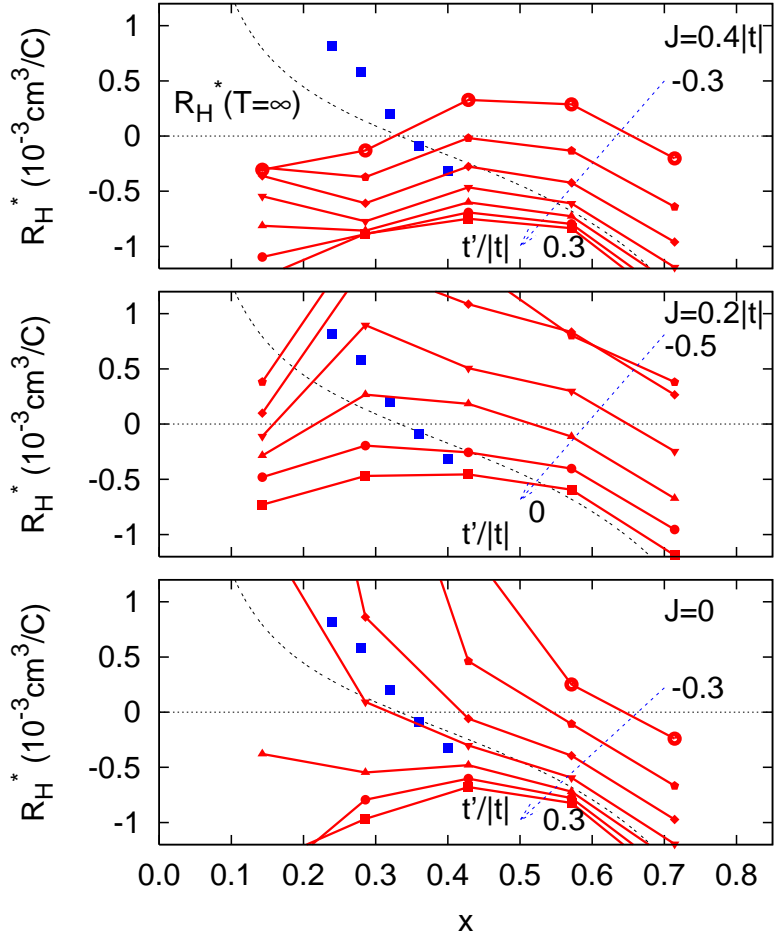}
  \caption{(color on-line) Hall coefficient in experimental units as function of doping for different values of $t'$.
The dots are from the experimental data in\cite{tsukada}, extrapolated to $T\rightarrow 0$. In the top panel,  
 $R_H^*(T=\infty)$ is the high-temperature limit for the case $t'=0$\cite{sss}. In these curves $J$ is varied from $0$ to $.4 |t|$. The bottom and top most  sets of curves correspond to $-0.3\leq t'/|t| \leq 0.3$ in steps of $0.1$ whereas the middle  one has  $ -.5 \leq t'/|t|\leq 0 $. The upper curves suggest that the number of zero crossings of the Hall constant is 3 for $0 \leq x \leq 1$, since  at small enough $x$, $R_H$ must show an upturn towards $+\infty$,  due to the Mott Hubbard insulating state at half filling.}
\label{Hall_coefficient}
\end{figure}
\end{center}

To gain a further understanding of the effect of $t'$ we introduce a finite value of $J=0.4 |t|$ and compare with the case of $J=0$. In Fig. \ref{J_depd} we show for several values of $t'$ the numerator of the high-frequency Hall-coefficient $R_H^*$. It is sufficient for our purposes to investigate this quantity, as it determines the  possible zero-crossings of the Hall-coefficient. The denominator contains $\langle \tau^{xx}\rangle $, which is a rather well-behaved quantity, varying only slightly in magnitude by introduction of a small second-neighbor hopping. It vanishes in the limits of $x\rightarrow 0$ and $x\rightarrow 1$ and ultimately leads to a divergence of $R_H^*$ in these two limiting cases. 

In  Fig.\ref{J_depd}, it is  instructive to begin with the case of $t'=0$ (lowest panel). Here, for $J=0$ we obtain the zero-crossing at $x=1/3$, similar to a prediction from high-temperature expansions\cite{sss}. Introducing a finite $J$  shifts the zero-crossing to lower dopings,  ($x \sim 0.15$). Here  $J$ acts as a source of antiferromagnetic correlations. Phenomenologically speaking, these antiferromagnetic correlations tend to resist a zero-crossing. In a sense, this is similar to the effect of the triangular lattice with a frustrated ($t'>0$) hopping amplitude\cite{curie_weiss}, where the zero-crossing is shifted to lower dopings. If the second-neighbor hop $t'>0 $ of the sign corresponding to an {\it electronically frustrated} system is now explicityly  introduced (left five panels of the figure), we find  an almost perfect alignment of the two curves of different $J$. Thus adding $t'>0$ has a similar effect to adding an antiferromagnetic $J$. This is quite consistent with our premise that this effect can be interpreted in terms of the so called
kinetic antiferromagnetism, or counter Nagaoka Thouless physics at play in frustrated triangular loops. \cite{counter_nagaoka}. On the other hand, if $t' <0 $  {i.e. a \it non-frustrated} sign is employed (right five panels of figure), the divergence between  the two curves becomes much more pronounced. We may loosely attribute this to  the ferromagnetic Nagaoka Thouless tendency towards a large fermi surface. Finite $J$ then dramatically destroys this state, especially near half-filling where its relevance is much stronger than close to the band-limit.

Thus an understanding of the sign of the Hall constant and its dependence on the sign of the hopping $t'$ seems to be closely linked to understanding the magnetic implications of the sign of $t'$. We emphasize that these trends refer to the tendencies of the correlated matter towards various kinds of magnetically ordered states, but do not invoke any actual broken symmetries. Hence these are a statement 
about underlying short ranged  correlations in the many body system.
\begin{flushleft}
\begin{figure}[h]
\includegraphics[width=13cm]{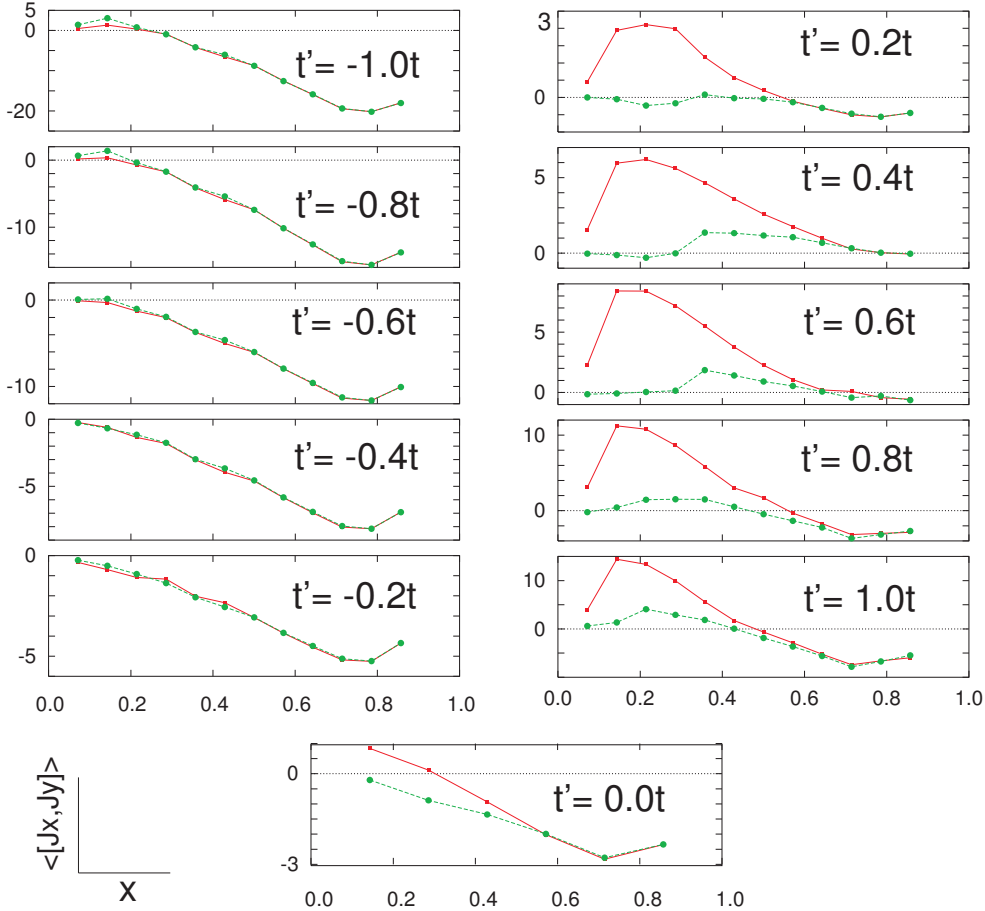}
  \caption{(color on-line) $t'-J$-dependence of  the (dimensionless) $\langle \Psi_0|[J_x,J_y]|\Psi_0\rangle $: red curves (green curves) correspond to  $J=0$ ($J=0.4|t|$). Note that for $t'/|t| >0$ a finite value of $J$ has little effect while for the opposite sign the effect is very pronounced. This is due to the observation that a positive $t'/|t|$ plays the same role as a positive $J$ from the  counter Nagaoka Thouless physics\cite{counter_nagaoka}.  }
\label{J_depd}
\end{figure}
\end{flushleft}

\subsection{Optical Sum Rule and Hall Number}
To evaluate the optical sum we make use of the relations Eqn. \ref{f_sum_rule} and \ref{x_optical}.  This  is shown in Fig. \ref{n_eff} for two values of J. We  observe  that this   quantity is   strikingly different  from the inverse Hall constant. The optics derived $x_{opt}$ follows roughly the  chemical doping $x$ and increases in magnitude as function of $t'$.  One noticable feature is that a naive linear extrapolation of the  small $x$ results misses the origin slightly: thus presumably there is a change in slope for smaller $x \leq .12$. It shows a maximum at intermediate dopings $x\approx 0.6$ as the trade-off for the stress-tensor between the available hole and electron carriers is optimized here. In particular,  $x_{opt}$ remains unaffected by the change in sign of the Hall constant when it occurs. 
\begin{center}
\begin{figure}[h]
\includegraphics[width=8.5cm]{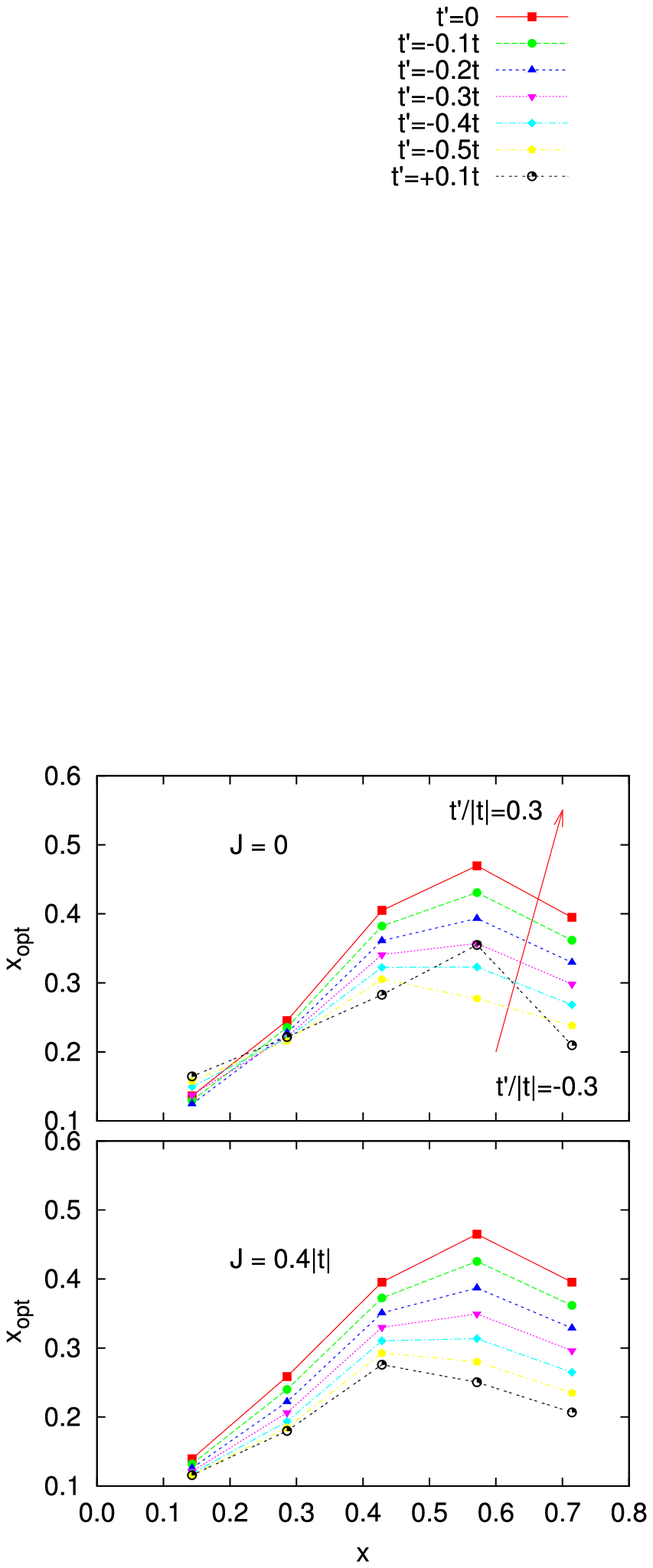}
  \caption{(color on-line) : Doping dependence of the effective particle number derived from optical sum rule, computed on the 14-site cluster shown in Fig.\ref{14_site_square}. The two sets of curves for $J=0, 0.4 |t|$ are qualitatively similar, except for $ t'/|t| \sim  -.3$, where a finite $J$ smooths out the sharp change at $J=0$.}
\label{n_eff}
\end{figure}
\end{center}
We now examine optimum doping, motivated by recent experimental results on the Hall-number in this range of doping\cite{balakirev}. Experimentally, the Hall number shows rather unusual nonlinear dependence on chemical doping $x$. To understand this behavior, we examine more closely the high-frequency limit $R_H^*$ near doping $x=0.15$. This corresponds to the introduction of two holes into finite systems of 14 and 15 sites. The case of a single hole is numerically  ill-behaved. Hence, for $x=2/L$ we study the dependence of $R_H^*$ on $t'/|t|$ and $J$ in a physically meaningful range of values. In Fig. \ref{JXJY} we present the numerator of Eqn. \ref{R_H_star} as function of $t'/|t|$ for several values of $J$ and the two systems studied. The figure shows that the dependence on $t'/|t|$ is rather pronounced, leading to a zero-crossing in the case of $J=0$. However, a small but finite value of $J$ tends to destroy the strong $t'$ dependence. 
\begin{center}
\begin{figure}[h]
\includegraphics[width=12cm]{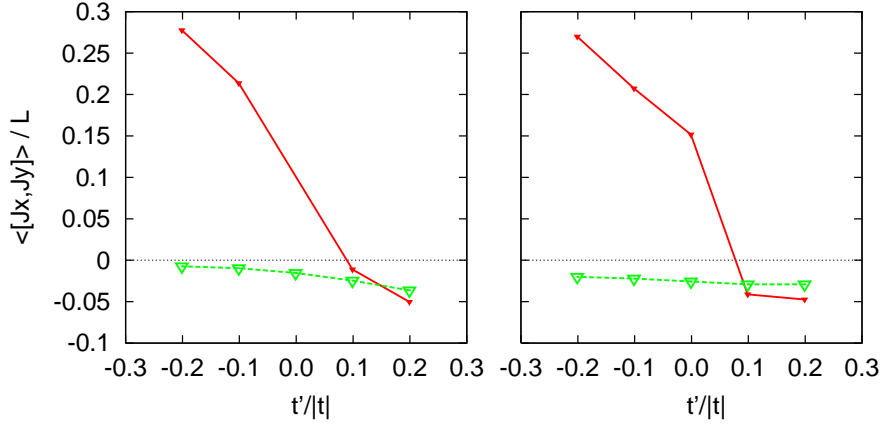}
  \caption{(color on-line) $\langle [J_x,J_y]\rangle$ (dimensionless units) as function of $ t'/|t|$ near optimal doping $x=0.14$ ({\bf l}) and $x=0.13$ ({\bf r}) computed on 14 and 15 site clusters, respectively. Red (green) curve is for $J=0$ ($J=0.4|t|$). Both clusters yields similar results: $\langle [J_x,J_y]\rangle$ is much more sensitive to $J$ at negative values of $t'/|t|$.  }
\label{JXJY}
\end{figure}
\end{center}
The particle number $x_{opt}$ obtained from the optical sum rule is shown in Fig. \ref{XOPT}. Its $t'/|t|$-dependence is much weaker than that of $\langle [J_x,J_y]\rangle$.
\begin{center}
\begin{figure}[h]
\includegraphics[width=12cm]{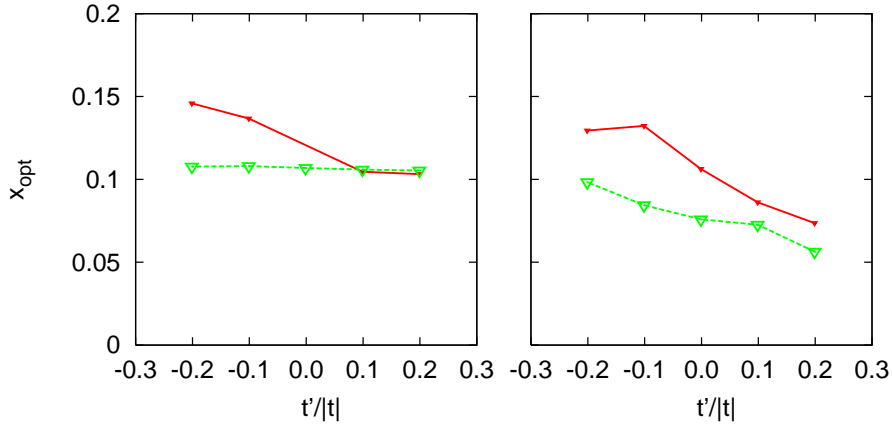}
  \caption{(color on-line) $x_{opt}$ as function of $t'/|t|$ near optimal doping $x=0.14$ ({\bf l}) and $x=0.13$ ({\bf r}) computed on 14 ({\bf l}) and 15 ({\bf r}) site clusters obtained by using Eqn. \ref{f_sum_rule}.  Red (green) curve is for $J=0$ ($J=0.4|t|$). Both clusters show similar trend: $x_{opt}$  weakly effected by $t'/|t|$, the effect is stronger for $J=0$. }
\label{XOPT}
\end{figure}
\end{center}
By combining Fig. \ref{JXJY} and \ref{XOPT} we obtain the inverse Hall-number, shown in Fig. \ref{XHALL}. The comparison of the $t'/|t|$ and $J$ dependence of this quantity with that of $x_{opt}$ makes two points very clear: (I) The optical particle number and the Hall number are fundamentally different objects in strongly-correlated systems. (II) The explanation of the experimentally measured nonlinear Hall-number\cite{balakirev} lies in a complicated interplay between the effect of finite - but probably small - $J$ and a non-zero value of $t'/|t|$ which allows an electron-like Hall-coefficient in the optimally doped regime.
\begin{center}
\begin{figure}[h]
\includegraphics[width=12cm]{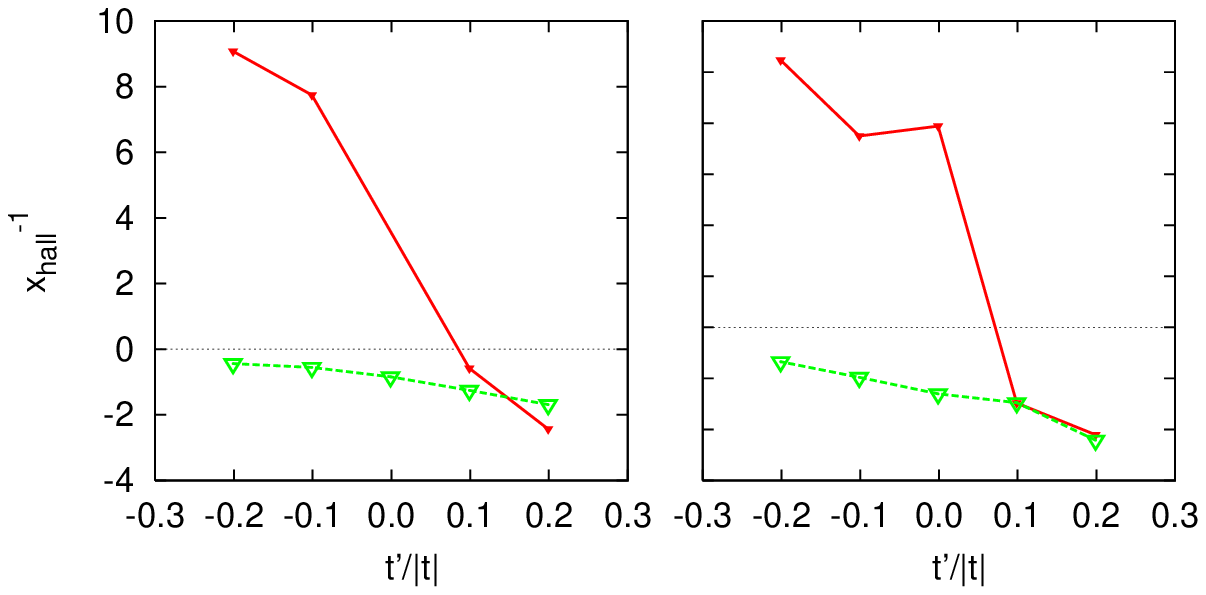}
  \caption{(color on-line)  $x_{hall}^{-1}$ as function of $t'/|t|$ near optimal doping $x=0.14$ ({\bf l}) and $x=0.13$ ({\bf r}) computed on 14 ({\bf l}) and 15 ({\bf r}) site clusters. The red (green) curve is for $J=0$ ($J=0.4|t|$) .  We see that  $x_{hall}$ is a sensitive function of $t'/|t| $ with a change of sign for $J=0$ near $t'/|t| = 0.1$.  }
\label{XHALL}
\end{figure}
\end{center}
Fig. \ref{x0_vs_t_prime} shows the doping value of the zero-crossing $x_{zc}$ of $R_H^*$ as function of 
$ t'/|t|$ for two different values of $J$ (compare Fig. \ref{J_depd}). The data in this plot (upper panel) is  for the hole-doped situation, and in the lower panel for the electron doped case using  the transformation \cite{kumar} 
\beq R_H(t,-t',n)= -R_H(t,t',2-n).\label{e_h}\eeq 
In the extreme  limits $ |t'/t| \rightarrow \infty$ the value of $x_{zc}$ approaches that of the case $t' =0$ since these two cases correspond to nearest-neighbor hopping on a bipartite square lattice, hence the sign is irrelevant in these limits. However  for $0.6\geq t'/|t| \geq 0.2$  the zero-crossing appears to disappear  in the ground state, within the limits of our calculation. This disappearance is rather independent of the value of $J$. 

\begin{center}
\begin{figure}[h]
\includegraphics[width=12cm]{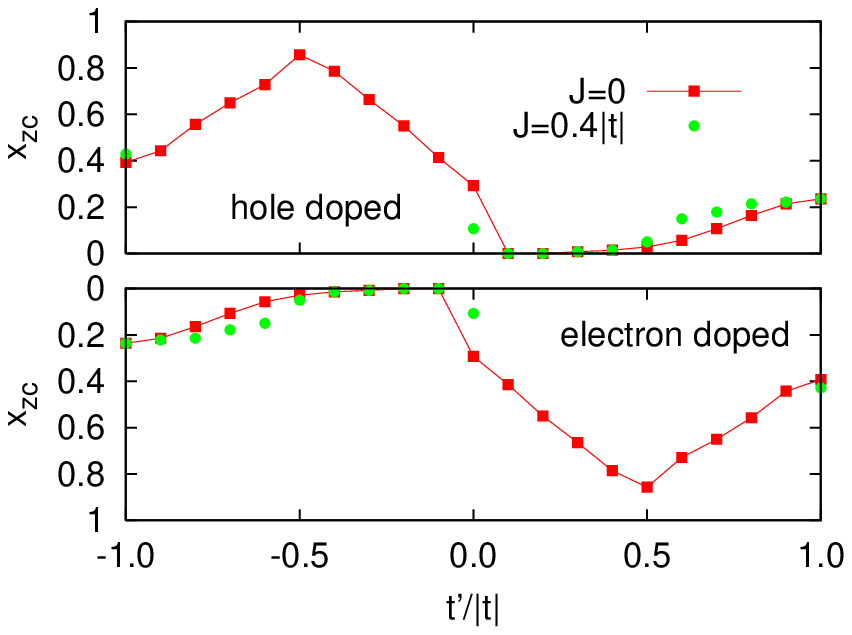}
  \caption{(color on-line)  The ground state Hall-coefficient zero-crossing doping $x_{zc}$ vs $ t'/|t|$ for two different values of $J$ for hole doping (top panel) and electron-doping (bottom panel), curves related by Eqn. \ref{e_h}. Our finite systems do not allow determination of $x_{zc}$ for all $t'/|t|$, we extrapolate doping dependence (Fig. \ref{J_depd}) to obtain estimates for $x_{zc}$.  }
\label{x0_vs_t_prime}
\end{figure}
\end{center}

\section{Conclusion}

 In the High Tc materials, spin fluctuation ideas by Kontani and others\cite{kontani},  lead to doping and frequency dependence of the Hall constant that are interesting. Also Kotliar and coworkers\cite{kotliar}  have studied the Hall constant using dynamical mean field theory ideas. In this work, we have studied a strong coupling model, namely the $t-t'-J$ model by using a combination of theoretical ideas and computation of the exact spectrum of the model for small clusters. The  problem addressed is that for classical metals the chemical doping $x$, the Hall number $x_{Hall}$ and particle number $x_{opt}$ derived from the optical sum rule agree well. However,  in strongly-correlated systems, they follow completely different ``renormalization paths''.  In the present study of the $t-t'-J$ model, we have extended previous studies to include the second neighbor hopping $t'$. This term plays a crucial role in determining the detailed behavior of the Hall constant. The Hall-number diverges at certain dopings and values of second neighbor hopping $t'$. Furthermore, it strongly depends on the value of the interaction strength $J$ for the case of positive (non-frustrated) $t'$.  This unusual result is understandable in terms of the concept of ``electronic frustration'',  a form of quantum interference.

 The inferred particle number $x_{opt}$  increases roughly with the chemical doping $x$ as the hole-number increases. We do see a signature of a different slope for very small $x\leq .12$. However, once the optimum trade-off between particle-density and carrier-freedom is reached, this quantity begins to decline and hence departs from the the value of $x$. Near optimal doping, we show that both $t'$ and $J$ significantly impact on the sign and magnitude of the Hall number. The optically derived hole number is much better behaved, i.e. its dependence on parameters is milder, and therefore seems a  safer
object to infer filling from. 

 In reconciling our numerical results with the recent experiments of Ref\cite{tsukada} on clean films of  LSCO, we concur with these authors that the data at $T\sim 300K$ is safer to compare with the present type of  theory, since the  $T$ sensitivity is out of our theoretical reach. The absolute values of the Hall constant for $x\geq .24$  found by them (their Fig.3) are roughly comparable to what we find, although we do need to vary the parameters more systematically for attempting an actual  fitting.    Their recognition that larger $x \geq .3 $ leads to an unbounded growth of the (negative) Hall constant is   important. It shows that the intrinsic behavior of data is in keeping with our ideas of Mott Hubbard physics versus uncorrelated band physics. This is  explained in Ref\cite{tsukada,sss}, where it is pointed out that  in   the limit $x\rightarrow 1$,  the Hall constant must be simply $R_H \sim - v/|q_e|(1-x)$, due to the proximity of the   band edge.

While our results are on quite small systems presently, they shed light on the questions arising from experiment, namely a variety of changes of sign and unusual magnitudes of the Hall constant in different High $T_c$ systems. This  study  also extends the insights of SSS Ref\cite{sss} and Stanescu and Phillips\cite{phillips} on a  Mott Hubbard theory of the Hall constant. Further detailed numerical studies could help produce systematic tables from which parameters could be inferred, and thus help in subclassifying the  High $T_c$ materials more precisely.

\begin{acknowledgments}

We gratefully acknowledge support from Grant No. NSF-DMR0408247 and DOE-BES DE-FG02-06ER46319. We thank F. Balakirev,  D. Basov and G. Blumberg, G. H. Gweon and H. Takagi for stimulating discussions.

\end{acknowledgments}
%\bibliography{bibl}

\begin{thebibliography}{99}
\bibitem{momentum} The fermi liquid relation\cite{nozieres}  $\frac{m}{m^*}= ( z_k  \frac{\partial \Sigma(k,E(k))}{\partial k_x})/(\partial \varepsilon_k/\partial k_x) $ implies that a vanishing of $z_k$ leads to a divergent effective mass, unless the momentum dependence is  divergent. Since the observed specific heat is not particularly enhanced in High $T_c$ systems, and further  $z_k$ is zero (or  very small), we must conclude that the momentum dependence of the self energy $\Sigma(k,\omega)$ is very severe.
\end{thebibliography}

\end{document}